\documentclass[a4paper,fleqn,usenatbib]{mnras}

\usepackage{mathptmx}

\usepackage[T1]{fontenc}
\usepackage{ae,aecompl}

\usepackage{graphicx}	
\usepackage{amsmath}	
\usepackage{amssymb}	

\usepackage{times}

\usepackage{url}

\newcommand{\teff}{$T_{\mathrm{eff}}$}
\newcommand{\muhz}{$\mu$Hz}
\newcommand{\numax}{$\nu_{\mathrm{max}}$}
\newcommand{\dnu}{$\Delta\nu$}

\newcommand{\kepler}{\textit{Kepler}}
\newcommand{\keplers}{\textit{Kepler's}}

\title[TESS asteroseismology of the Kepler red giants]{TESS asteroseismology of the Kepler red giants}

\author[D. Stello et al.]{
Dennis~Stello$^{1,2,3}$ 
Nicholas Saunders$^{4}$          
Sam Grunblatt$^{5,6}$        
Marc Hon$^{1,4}$             
Claudia Reyes$^{1}$\and          
Daniel~Huber$^{4}$           
Timothy~R.~Bedding$^{2,3}$   
Yvonne Elsworth$^{7}$        
Rafael~A.~Garc\'ia$^{8}$          
Saskia Hekker$^{9,10,3}$\and     
Thomas Kallinger$^{11}$      
Savita Mathur$^{12,13}$       
Benoit Mosser$^{14}$          
Marc~H.~Pinsonneault$^{15}$   
\\
$^{1}$School of Physics, University of New South Wales, NSW 2052, Australia\\
$^{2}$Sydney Institute for Astronomy (SIfA), School of Physics, University of Sydney, NSW 2006, Australia\\
$^{3}$Stellar Astrophysics Centre, Department of Physics and Astronomy, Aarhus University, DK-8000 Aarhus C, Denmark\\
$^{4}$Institute for Astronomy, University of Hawai`i, 2680 Woodlawn Drive, Honolulu, HI 96822, USA\\
$^{5}$American Museum of Natural History, 200 Central Park West, Manhattan, NY 10024, USA\\
$^{6}$Center for Computational Astrophysics, Flatiron Institute, 162 5thAvenue, Manhattan, NY 10010, USA\\
$^{7}$School of Physics and Astronomy, University of Birmingham, B15 2TT, UK\\
$^{8}$AIM, CEA, CNRS, Universit\'e Paris-Saclay, Universit\'e Paris Diderot, Sorbonne Paris Cit\'e, F-91191 Gif-sur-Yvette, France\\
$^{9}$Center for Astronomy (ZAH/LSW), Heidelberg University, Königstuhl 12, 69117 Heidelberg, Germany\\
$^{10}$Heidelberg Institute for Theoretical Studies (HITS) gGmbH, Schloss-Wolfsbrunnenweg 35, 69118 Heidelberg,Germany\\
$^{11}$Institute of Astrophysics, University of Vienna, 1180 Vienna, Austria\\
$^{12}$Instituto de Astrofisica de Canarias, E-38200 La Laguna, Tenerife, Spain\\
$^{13}$Universidad de La Laguna (ULL), Departamento de Astrofisica,E-38206 La Laguna, Tenerife, Spain\\
$^{14}$LESIA, Observatoire de Paris, Universit\'e PSL, CNRS, SorbonneUniversit\'e, Universit\'e de Paris, 92195 Meudon, France\\
$^{15}$Department of Astronomy, The Ohio State University, Columbus, OH 43210, USA
}

\date{Accepted XXX. Received YYY; in original form ZZZ}

\pubyear{2020}

\begin{document}
\label{firstpage}
\pagerange{\pageref{firstpage}--\pageref{lastpage}}
\maketitle

\begin{abstract}
Red giant asteroseismology can provide valuable information for studying the Galaxy as demonstrated by space missions like CoRoT and \kepler.  However, previous observations have been limited to small data sets and fields-of-view. The TESS mission provides far larger samples and, for the first time, the opportunity to perform asteroseimic inference from full-frame images full-sky, instead of narrow fields and pre-selected targets.  Here, we seek to detect oscillations in TESS data of the red giants in the \kepler\ field using the 4-yr \kepler\ results as benchmark.  Because we use 1-2 sectors of observation, our results are representative of the typical scenario from TESS data.  We detect clear oscillations in $\sim$3000 stars with another $\sim$1000 borderline (low S/N) cases.  In comparison, best-case predictions suggests $\sim$4500 detectable oscillating giants.  Of the clear detections, we measure \dnu\ in 570 stars, meaning a $\sim$20\% \dnu\ yield (14\% for one sector and 26\% for two sectors). These yields imply that typical (1-2 sector) TESS data will result in significant detection biases.  Hence, to boost the number of stars, one might need to use only \numax\ as the seismic input for stellar property estimation.  However, we find little bias in the seismic measurements and typical scatter is about 5-6\% in \numax\ and 2-3\% in \dnu.  These values, coupled with typical uncertainties in parallax, \teff, and [Fe/H] in a grid-based approach, would provide internal uncertainties of 3\% in inferred stellar radius, 6\% in mass and 20\% in age for low-luminosity giant stars.  Finally, we find red giant seismology is not significantly affected by seismic signal confusion from blending for stars with Tmag $\lesssim 12.5$.    
\end{abstract}




\section{Introduction}
The space-based asteroseismic revolution of red giant stars \citep{Ridder09} spawned the realisation that oscillating giants would provide powerful ways to study the Milky Way \citep{Miglio09}.  The initial attempts of this asteroseismically-informed Galactic archaeology were made with CoRoT \citep[e.g.][]{Miglio13,Anders17} and later with \kepler\ \citep[e.g.][]{Sharma16,Casagrande16,Silva18}.  However, it soon became clear that the small sky coverage and the complex, and to some degree undocumented, target selection function would limit the use of these particular data sets within this line of research.  Fortunately, \keplers\  K2 mission \citep{Howell14} gave birth to the K2 Galactic Archaeology Program designed to support studies of the Milky Way along the ecliptic, with stars probing many different parts of the Galaxy and following a simple reproducible selection function \citep{Stello15,Sharma22}.  Although seismic data have been released for all campaigns of the K2 Galactic Archaeology Program \citep{Stello17,Zinn20,Zinn21b}, the scientific fruits of this rich data set have only just started to be harvested \citep{Sharma19,Rendle19,Khan19,Sharma21}.

The launch of NASA's TESS mission opened the first opportunity to detect oscillations in red giants over the full sky \citep{Ricker15,Campante16}, with its initial 2-year mission covering first the southern ecliptic hemisphere, followed by the northern hemisphere.  The potential to study large stellar populations in the Milky Way with TESS is therefore significant.  In an early attempt to quantify the asteroseismic performance of TESS in this context, \citet{Silva20} used TASOC\footnote{TESS Asteroseismic Science Operations Center: www.tasoc.dk} `FastTrack' data of 25 bright red giants ($V\simeq6$) from the first two sectors of TESS's southern hemisphere observations. They found all the giants in their sample showed oscillations, confirming the expected TESS performance.  When combining the seismology from TESS with parallaxes from Gaia DR2 \citep{Gaiadr2}, they found the precision on the inferred stellar radii, masses, and ages from grid modelling was similar to that obtained from 4-year \kepler\ data.  This showed that the smaller aperture and shorter observation time span by TESS (leading to less precise seismic measurements) is compensated in the grid modelling by the targets being brighter and closer (more photons and more precise parallaxes) compared to the typical \kepler\ targets.
Later, \citet{Mackereth21} used a full-year (13 sectors) of TESS southern-continuous-viewing-zone data, covering about 450 square-degrees, to infer the potential for red giant asteroseismology with TESS across its full-sky view.  They estimated $\sim$300,000 giants would show oscillations across the sky.

During its second year, TESS covered the \kepler\ field in Sectors 14 (fully) and 15 (partly). This provided an interesting opportunity to test the TESS performance in more detail on a large sample of well-studied red giants.  Despite the limitations of the \kepler\ data for Galactic archaeology studies, the mission provides the best quality data for red giant seismology on individual stars.  As such, \kepler\ still is the benchmark for red giant seismology.  The nearly continuous observations for four years, stable environment far from the Earth, and relatively large aperture means that \kepler-based results probably will remain the ultimate `ground truth' for the foreseeable future.
In addition to testing the TESS performance, the TESS observations of the \kepler\ red giants also gives us an important way to verify whether our seismic measurements are consistent with the `true' values, as we move toward analysing all TESS data fully automatically in future. 

In this paper, we use the \kepler\ results on red giants to study how well we can measure the oscillations from TESS data of all giants in the \kepler\ field brighter than \texttt{Kp}$=13$.  Particularly, we want to (1) investigate how the intrinsic limitations of TESS (such as small aperture and short observation time) affects the completeness
of the seismic stellar population from TESS,
(2) study if the uncertainties on the seismic observables \numax\ and \dnu\ are representative of the true uncertainties, (3) estimate the yield of stars with reliable \dnu\ measurements as opposed to only \numax, (4) see if there is any bias in \numax\ and \dnu\ relative to the \kepler\ results, and finally, (5) provide a rough estimate of the radius, mass, and age precision one can expect from the one to two sectors of TESS observations.

\section{Target selection and light curve creation}\label{observations}
We selected the 8668 stars brighter than \texttt{Kp}$=13$ in the catalogue of 16000 \kepler\ red giants with detected oscillations by \citet{Yu18}.  These stars were all observed with \keplers\ 30-minute cadence and have a measurement of the frequency of maximum acoustic power, \numax, and of the frequency separation between radial overtone modes, \dnu.

To cover as many stars as possible, we used the TESS Full Frame Images taken at 30-minute cadence as our data source.
We followed the approach of \citet{Saunders21}\footnote{https://github.com/nksaunders/giants}, which we summarise here.  First, we retrieved data from the Mikulski Archive for Space Telescopes (MAST) using \textsf{TESScut} \citep{Brasseur19}
to download 11x11 pixel cutouts around each target and then apply the following methodology to remove the scattered light background from the TESS Full Frame Image observations. Our pipeline uses the {\sf RegressionCorrector} framework in the {\sf lightkurve} Python package \citep{Lightkurve18}.  Using the cutout target pixel files, we created a design matrix with column vectors populated by the flux light curves of pixels outside a threshold aperture mask, avoiding pixels that contain flux from the target to ensure our noise model did not fit out the desired signal.  We then performed Principal Component Analysis on the columns of the design matrix to find ten principal components to use in our model.  To produce our final noise model, we set up a generalized least-squares problem to find optimal coefficients for each of the components in our design matrix, and generated a model as a linear combination of the column vectors.  We produced an uncorrected light curve by performing simple aperture photometry on the cutout target pixel file using the inverse of the aperture mask used to select regressors.  Our final light curves were produced by subtracting the noise model from the uncorrected light curves. 

\begin{figure*}
\includegraphics[width=10.0cm,angle=-90]{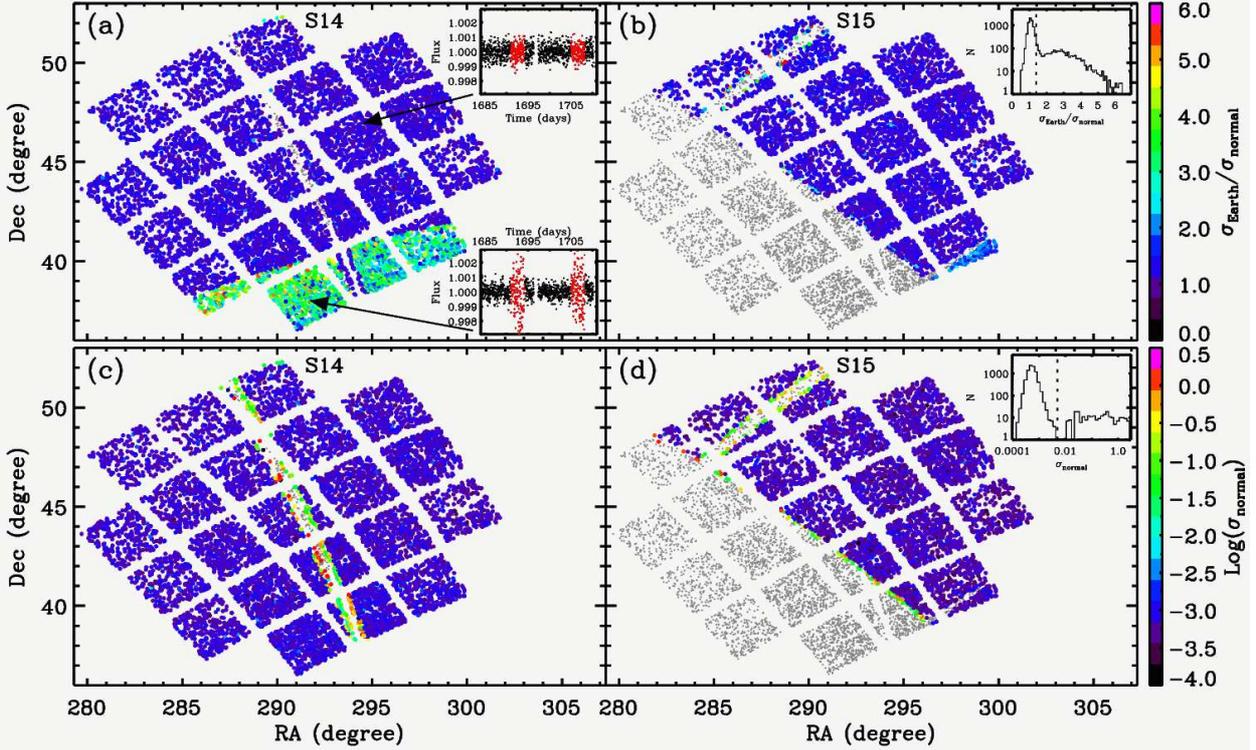}
\caption{Sky coverage of our targets (grey points). (a) Observed stars in TESS Sector 14 are colour-coded by $\sigma_\mathrm{Earth}/\sigma_\mathrm{normal}$. The insets show two example light curves, with the time stamps used to calculate the scatter colour-coded by red ($\sigma_\mathrm{Earth}$) and black ($\sigma_\mathrm{normal}$) points. (b) Same as (a) but for Sector 15. The inset shows the $\sigma_\mathrm{Earth}/\sigma_\mathrm{normal}$ distribution and the cut-off value of 1.4.  (c) Same as (a) but colour-coding showing log($\sigma_\mathrm{normal}$). (d) Same as (c) but for Sector 15. The inset shows the $\sigma_\mathrm{normal}$ distribution and the cut-off value of 0.005.
\label{fov}} 
\end{figure*} 
Almost all the selected stars were observed in Sector 14 (8576 stars) and about half were observed in Sector 15 (4909 stars).  We concatenated the light curves of those observed in both sectors (4817 stars).  We then followed the data processing previously applied to K2 data by \citet{Stello15,Stello17}, which included a 4-day wide boxcar high-pass filter (meaning a cut-off frequency of about 3\muhz\ in the frequency domain) and filling gaps below 1.5 hours in length using linear interpolation.

For each sector we identified the time stamp segments (spacecraft orbital phases) for which the light curves were potentially affected by Earth shine and subsequently removed affected stars \footnote{Although one could potentially salvage affected stars by removing only the affected time segments we opted not to do so for our purpose.}. Affected stars were defined as those with a light curve standard deviation in their potential Earth shine segments, $\sigma_\mathrm{Earth}$, above 40\% of their unaffected segments, $\sigma_\mathrm{normal}$. We removed 2307 stars in this process.

Figure~\ref{fov}a shows the sky coverage of our targets for Sector 14, revealing the footprint of the \kepler\ field of view.  The colour-code of each observed star represent $\sigma_\mathrm{Earth}/\sigma_\mathrm{normal}$. The part of the field affected by Earth shine (bright coloured dots) corresponds to TESS camera 1.  The two insets show example light curves with the segments potentially affected by Earth shine highlighted in red.  In Sector 15 the Earth shine issue is clearly less severe, only affecting the lower-right corner, as seen in  Figure~\ref{fov}b. The inset in this figure shows the $\sigma_\mathrm{Earth}/\sigma_\mathrm{normal}$ distribution and the cut-off (dashed line) used to remove affected stars.

\begin{figure}
\includegraphics[width=\columnwidth]{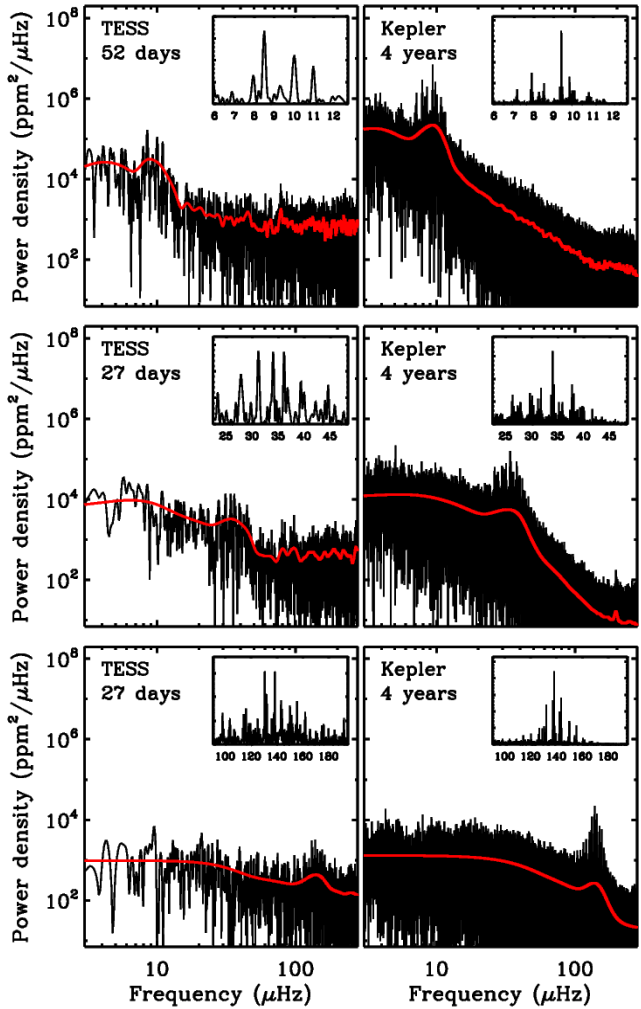}
\caption{Example power spectra from TESS (left) and \kepler\ (right) for three stars ranging from low to high \numax\ values. The red curves show the smoothed spectra using the same smoothing for TESS and \kepler.
\label{pow_spec_examples}} 
\end{figure}

We also found and removed an additional 196 stars that showed orders-of-magnitude higher noise than the rest of the sample, with a standard deviation $\sigma_\mathrm{normal} > 0.005$. All turned out to lie close to the TESS CCD edges (Figure~\ref{fov}c-d).  For the remaining 6165 stars we calculated the Fourier transform (power spectrum) for subsequent oscillation analysis.
In Figure~\ref{pow_spec_examples} (left panels) we show a representative set of the power spectra from TESS.  In the right panels we illustrate the corresponding \kepler\ data, which can be regarded as providing the ground truth benchmark measurements in this investigation. We note that amplitude calibration between TESS and \kepler\ is still uncertain (Lund et al. in prep) but will not affect the results presented here. 
Thirty stars in our sample also had 2-minute cadence TESS data, and hence an existing SPOC light curve on MAST, and comparison of those power spectra with ours showed on average similar power levels across all frequencies, although with some star-to-star variation.

\section{Detection of oscillations}
For Galactic archaeology in particular, we would like our seismic detection algorithms to provide complete and pure samples, meaning we detect all possible detections without introducing any false positives.  \citet{Stello17} demonstrated that visual inspection of power spectra provided a robust determination of which stars showed oscillations (high completeness and high purity), despite being subjective and time consuming. Based on this, and previous work by \citet{Hekker11} and \citet{Hekker12}, \citet{Yu18} used visual inspection to classify detections and non-detection for their sample of 15,000 \kepler\ red giants, now regarded a gold standard dataset from the \kepler\ red giants \citep[e.g.][]{Mackereth21}. 
To eliminate the shortcomings of performing visual inspection manually, \citet{Hon18a} trained an image-recognition artificial neural network on such visual classification, which was shown to be very efficient on \kepler\ data \citep{Hon19}.  However, this network has not yet been trained to provide both pure and complete sets of detections from actual TESS data.  We therefore followed the approach by \citet{Stello17} to manually classify our relelatively small sample of TESS stars into three detection categories: `Yes', `Maybe', and `No'.  These results helped inform our subsequent results when we came to assess how well we could measure the seismic, as well as fundamental global, properties of the stars.  

\begin{figure*}
\includegraphics[width=17cm]{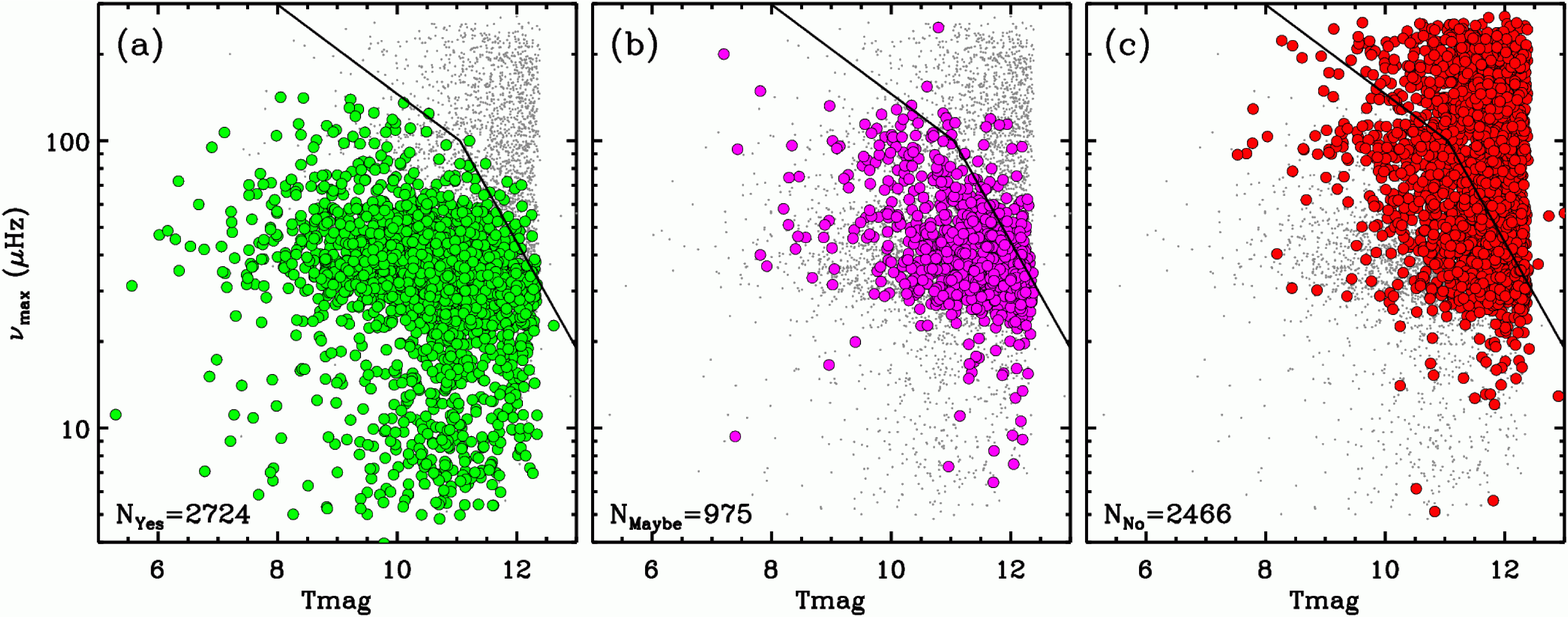}
\caption{\kepler\ values of \numax\ from \citet{Yu18} versus TESS magnitude from \citet{Stassun19} of all stars seismically analysed here (grey dots). In each panel they are colour-coded according to our detection of oscillations in TESS data: (a): Yes (green), (b): Maybe (magenta), (c): No (red).  The black line shows the predicted detection threshold for TESS.
\label{numax_vs_tmag}} 
\end{figure*}
Figure~\ref{numax_vs_tmag} shows the entire sample of stars, with the detection of oscillations by TESS indicated by colour.  We see that the detections (Figure~\ref{numax_vs_tmag}a green) follow a similar threshold trend in the upper right corner to that predicted using the formalism in \citet{Chaplin11,Schofield19} (black line).  For the predictions we ignored blending and systematic noise and different to the approach by \citet{Schofield19}, we used Gaia-based radii directly from the TESS Input Catalog \citep{Stassun19} and used TESS magnitudes in place of Johnson I-band.  Fainter and intrinsically less luminous stars (lower amplitude and larger \numax), have a signal-to-noise ratio too low to detect the oscillations.
Extrapolating the threshold line towards the most luminous giants with \numax\ $\sim 5-10\,$\muhz, suggests that TESS would probably be able to detect oscillations in stars as faint as Tmag$\sim 14$ at least for the most luminous stars. 
As expected, most of the `Maybe' detections (Figure~\ref{numax_vs_tmag}b magenta) are close to the detection threshold; they truly are borderline cases.  Many of them are situated in the red clump region around \numax\ $\sim 30-100\,$\muhz, which often provide lower and wider oscillation power excess detections \citep[e.g.][]{Mosser12a,Yu18}.  
While most non-detections (Figure~\ref{numax_vs_tmag}c red) are above the predicted threshold line, as expected, many fall well within the predicted `detection' region below the line.  Based on spot checks, many of them show either unusually strong low-frequency variation (regular or irregular, indicative of binarity or instrumental/photometric issues) or significantly different noise levels between the two observing sectors.  This strong overlap between detections and non-detections in \numax-Tmag space is only seen in the observations. The detection predictions show very little overlap if plotted in the \numax-Tmag diagram.  This is a result of ignoring any systematics, demonstrating that the predictions represent the ideal scenario (single well-isolated stars and a perfectly performing instrument and photometric extraction).  With this in mind, we count the number of stars with a predicted detection probability larger than 99\% to be about 4500.  Hence, the observed yield relative to this optimistic scenario is $\sim 60$\% for clear detections (2724 stars), and $\sim 80$\% if the stars marked `Maybe' are also counted as genuine detections (975 stars).  
We note that these yields are not like-for-like comparable to those of \citet{Mackereth21} because our studies are complementary.  Differences between the two studies include: (1) the length and filtering of the time series, (2) the target selection, and (3) the definitions for what constitutes a detection.

\begin{figure*}
\includegraphics[width=17cm]{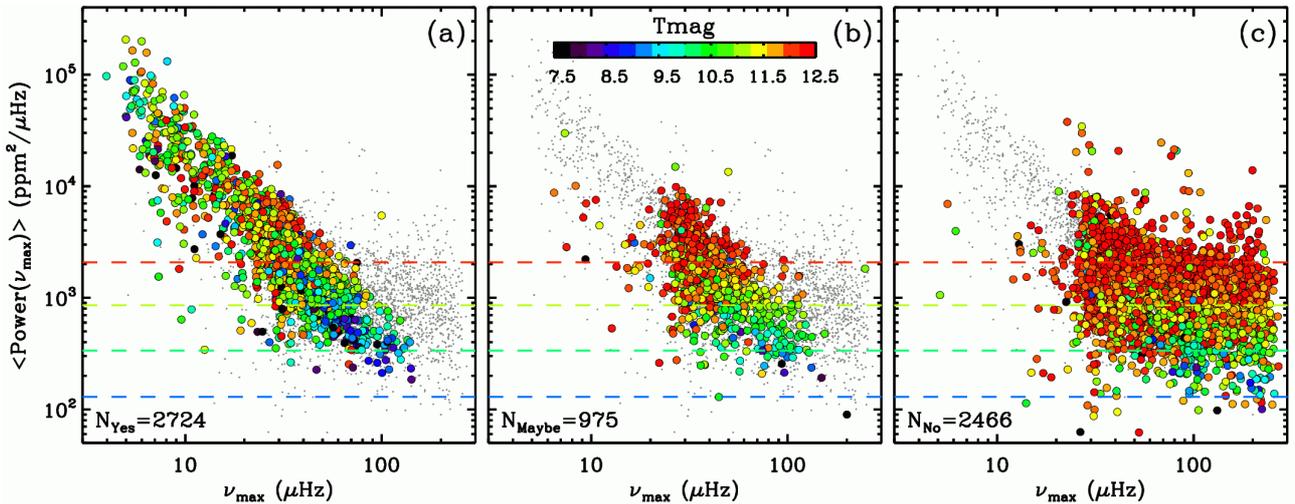}
\caption{Average power in the TESS data around \numax\ from \citet{Yu18} versus \numax\ of all the seismically analysed stars (grey dots). The colour-highlighted stars are separated in three panels according to their detection classification like in Figure\ref{numax_vs_tmag} (a: Yes; b: Maybe; c: No), but with the colour-coding showing Tmag. The dashed lines show the white noise levels according to Eq. 11 in \citet{Campante16} for Tmag = 9, 10, 11, and 12.   
\label{pow_vs_numax}} 
\end{figure*}
To further verify whether our detections follow expectations, Figure~\ref{pow_vs_numax} shows the average power in the TESS data as a function of \kepler\ \numax, measured in a 0.4\%\numax-wide window around the \kepler\ \numax.  The clear detections (Figure~\ref{pow_vs_numax}a) show a relatively tight power law relation with a sharp upper limit at fixed \numax, as seen in previous ensemble results \citep[e.g.][]{Yu18}, demonstrating that the power spectrum is dominated by oscillation power at \numax.  This is further supported by the power measured for a given star typically being much larger than the predicted white noise for its brightness (dots fall above dashed lines of the same colour).  Most of the `Maybe' detections (Figure~\ref{pow_vs_numax}b) also seem to follow the power law relation and power levels being higher than the predicted white noise, suggesting that they are mostly genuine detections.  The non-detections, however, mostly follow a flat and quite broad distribution (at fixed \numax), with many stars falling near and even below the predicted noise, which shows the power spectra are dominated by noise.  It is evident, however, that towards low \numax, some non-detections start to follow the steep power law of the detections, suggesting that some of these stars could possibly show hints of oscillation power.

\section{Seismic measurements}\label{seismic}
In the next step, we analysed the level of precision and accuracy in \numax\ and \dnu\ from the TESS data by benchmarking our results against the 4yr-based \kepler\ results by \citet{Yu18}.  The assumption is that the \kepler\ results can be regarded as the ground truth, with negligible uncertainty relative to that of the TESS measurements.  To make a like-for-like comparison, we followed the approach by \citet{Yu18} to extract \numax\ and \dnu\ using the so-called SYD pipeline by \citet{Huber09}, with improvements detailed in \citet{Huber11a} and \citet{Yu18}.  Here, we only looked at stars deemed clear detections in the previous section.

\begin{figure}
\includegraphics[width=\columnwidth]{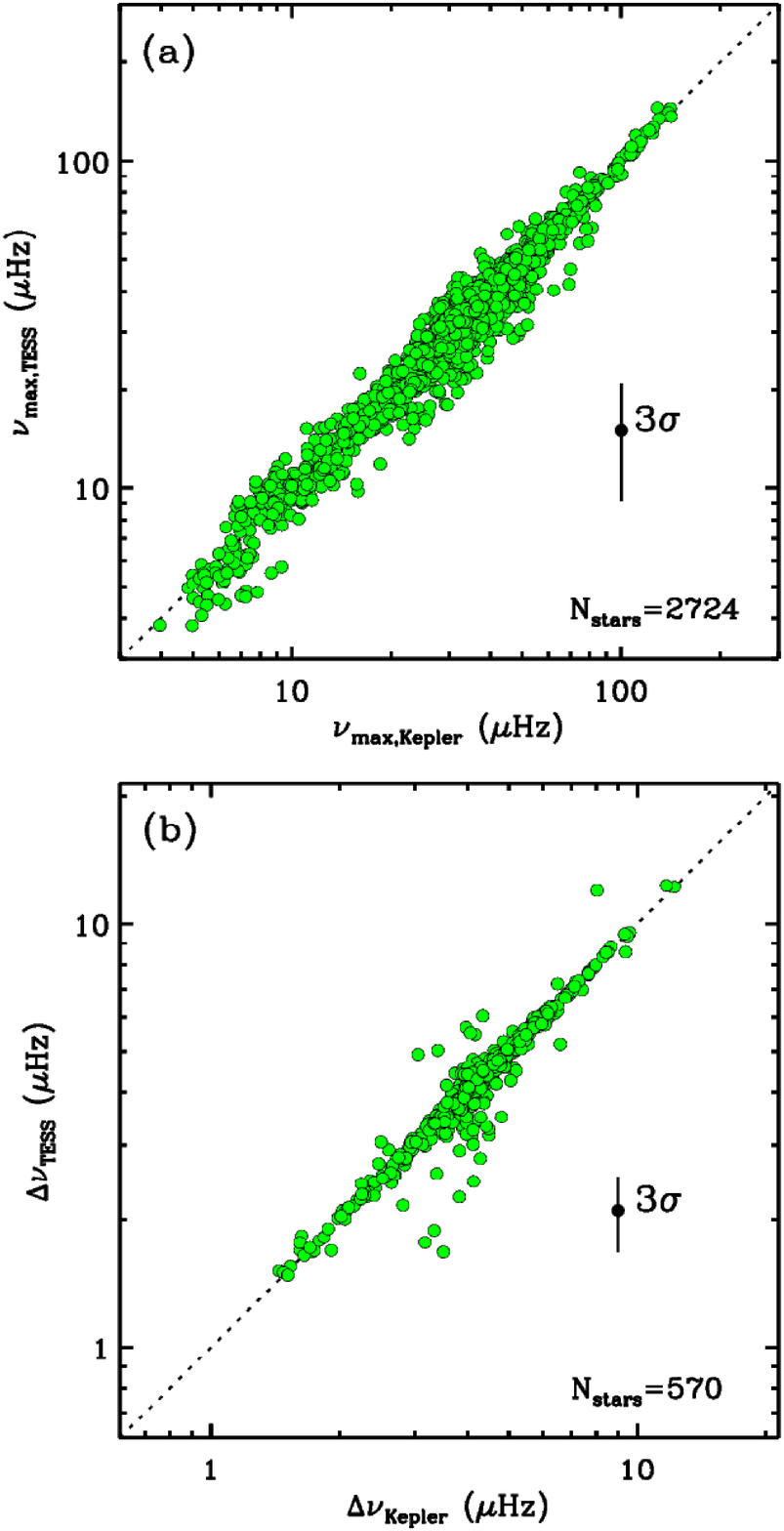}
\caption{TESS versus \kepler\ results for both \numax\ (a) and \dnu\ (b). Only stars with confirmed oscillations are shown in panel (a), while panel (b) shows only the subset that also have \dnu\ deemed reliable using our neural network vetter \citep{Reyes22}. The outliers have large quoted uncertainties.  The $3\sigma$ error bar represents the median uncertainties of the TESS data (the error bars for \kepler\ are too small to be visible).  
\label{tess_vs_kepler_numax_dnu}} 
\end{figure}
The direct comparison between the TESS and \kepler\ results is shown in Figure~\ref{tess_vs_kepler_numax_dnu}a for \numax\ and Figure~\ref{tess_vs_kepler_numax_dnu}b for \dnu.  The deviations from the dashed 1-to-1 line are completely dominated by the uncertainty in the TESS measurements (see representative $3\sigma$ error bars for TESS; \kepler\ error bars are too small to see).  The tight correlation in Figure~\ref{tess_vs_kepler_numax_dnu}a confirms that our detections  with TESS are robust.  A similar plot of the `Maybe' cases also reveals a tight relation, further supporting that most are genuine detections, while the `No' detections show an extremely large scatter indicative of random numbers. Almost all the outliers seen in Figure~\ref{tess_vs_kepler_numax_dnu}b have reported TESS uncertainties above 10\%. 

It is evident from Figure~\ref{tess_vs_kepler_numax_dnu}a that 1-2 sectors of TESS data will provide relatively few seismic detections of low-luminosity red giant branch stars (\numax\  $\gtrsim 100\,$\muhz) and of highly luminous giants (\numax\ $\lesssim 5\,$\muhz), with the bulk of detections being in the helium-core burning red clump stars (\numax\ $\sim 30-40\,$\muhz) (see also Figure~\ref{numax_vs_tmag}).
Unfortunately, red clump stars are typically the most difficult when it comes to extracting \dnu\ reliably from short time series, as evident from the larger spread in the red clump region of Figure~\ref{tess_vs_kepler_numax_dnu}b (\numax\ $\sim$ 3-4\muhz).

We know from previous careful visual inspection of K2 results, which covered $\sim80\,$days, that only about 50\% of the stars with oscillation power excess (a \numax\ detection) also provided reliable \dnu\ measurements \citep{Stello17}.  With one or two sectors of TESS data (27 days or 54 days) we would therefore expect somewhat lower yields.  To verify which stars had reliable \dnu\ detections, we used an improved version of the artificial neural network by \citet{Zinn20}, trained on one- and two-sector-long K2 data sets \citep{Reyes22}. 
We found that 570 stars showed reliable \dnu\ detections; hence an overall yield of 20\%.  In Table~\ref{tab1} we quantify the \dnu\ yields for different samples of stars and show how it depends on having one or two sectors of data.
\begin{table}
{\footnotesize
\centering
\caption{\dnu\ yields. \label{tab1}}
\begin{tabular}{lcc}
\hline
 Sample  & 1 sector   & 2 sectors  \\
\hline                                                                                                                              
Full     &  14\%      &     26\%  \\
RGB/AGB  &  20\%      &     48\%  \\
RC$^*$   &  12\%      &     19\%  \\
\hline
\multicolumn{3}{l}{*: Red clump (RC) star identifications are from \citet{Hon18b}.}\\
\end{tabular}
}
\end{table}
In addition to the shorter observations by TESS compared to K2, one reason why these yields are lower than for K2, could be that the lower signal-to-noise ratio in the TESS data (compared to K2), excludes predominantly low-luminosity red giant branch stars, which typically would provide a high fraction of \dnu\ detections due to their well-resolved simple frequency patterns \citep{Bedding10}.

\begin{figure}
\includegraphics[width=8.8cm]{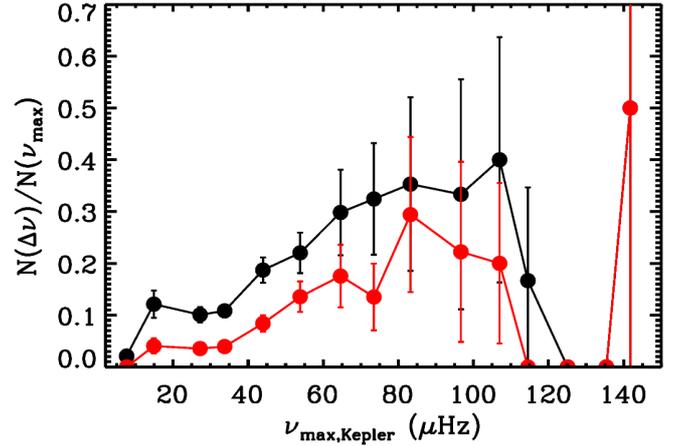}
\caption{Fraction of stars with a \dnu\ measurement to better than 3\% (black curve) and 1\% (red curve). 
\label{dnu_numax_ratio}} 
\end{figure}

In Figure~\ref{dnu_numax_ratio} we show the overall fraction of stars with \dnu\ measurements within 3\% and 1\% of the \kepler\ values as function of \numax\ to further demonstrate where the most and best results are expected.
In combination, Table~\ref{tab1}, and Figures~\ref{numax_vs_tmag}, \ref{tess_vs_kepler_numax_dnu}, and \ref{dnu_numax_ratio} imply that all regions of the parameter space (be it seismic or in brightness), and hence stellar evolutionary stage, are affected by detection bias. This clearly needs to be taken into account when assessing the completeness of the seismic samples for the purpose of population studies.

\begin{figure}
\includegraphics[width=8.8cm]{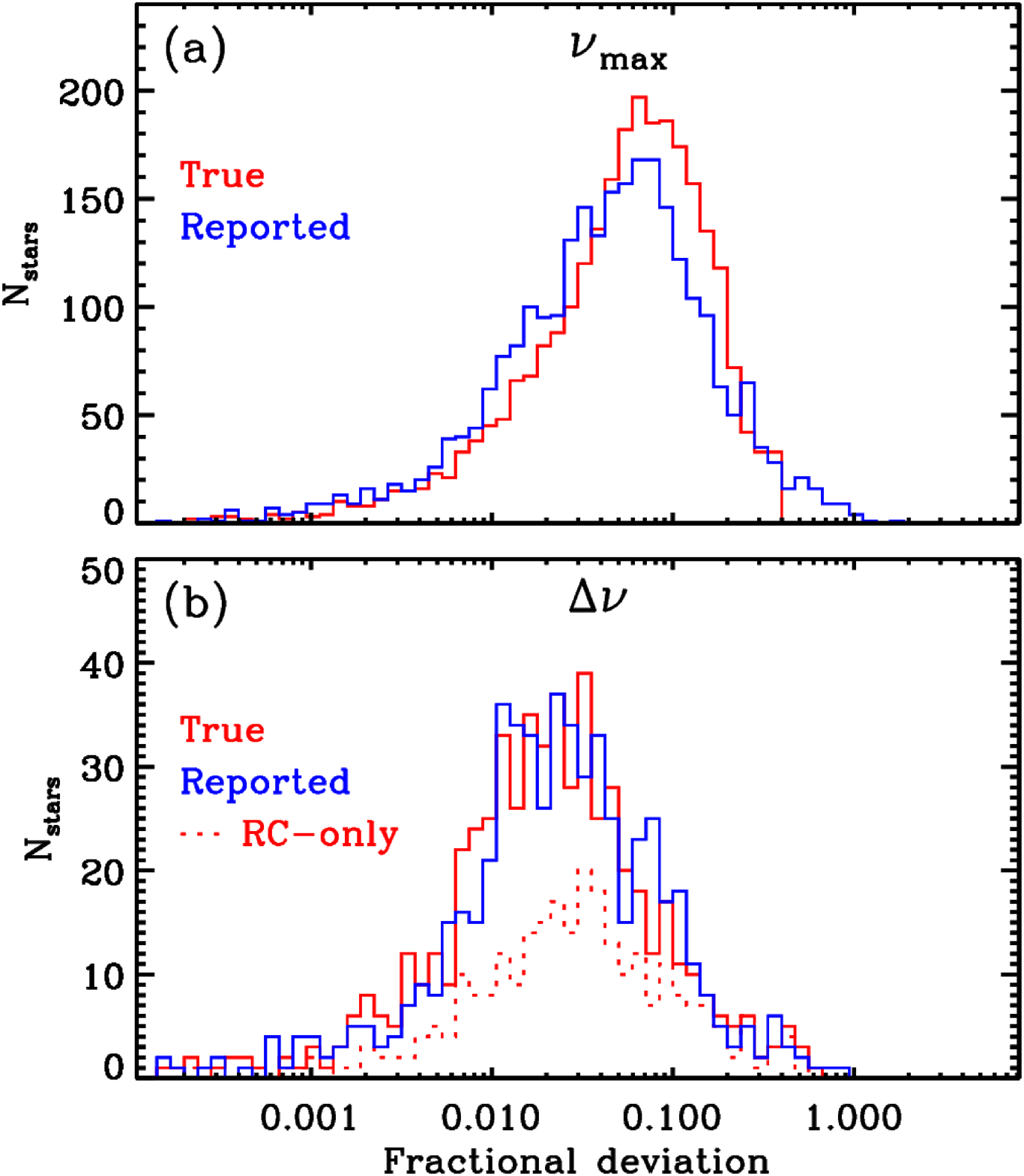}
\caption{Deviations of TESS results for both \numax\ (a) and \dnu\ (b). `True' deviations are |'TESS'$-$'\kepler'|/'\kepler'. `Reported' deviations are random extracts from N(0,$\sigma_\mathrm{'TESS'}$/'TESS') distributions.  
\label{true_vs_claimed_uncertainty}} 
\end{figure}
We now turn to the measurement uncertainties. The red histogram in Figure~\ref{true_vs_claimed_uncertainty}a shows the fractional deviation of the TESS \numax\ from the \kepler\ result (|\numax$_\mathrm{TESS}-$\numax$_\mathit{Kepler}|/$\numax$_\mathit{Kepler}$).  This deviation from the `true' value allows us to check if the reported uncertainties from the SYD pipeline are robust across the ensemble as a whole; in other words, whether they are representative of the true measurement uncertainties.  The blue curve in Figure~\ref{true_vs_claimed_uncertainty}a shows the deviation one would expect from the reported uncertainties. We derived each deviation by taking a random extract from a Gaussian distribution with a width $2\sqrt{2\ln2}$ times the reported uncertainty for each star\footnote{Adding the measurement uncertainty from the \kepler\ result to the width of the Gaussian did not significantly change the final distribution shown in the figure.}.  The distributions have similar shapes, although it seems the reported \numax\ uncertainties are on average underestimated by about 10-30\%.  

Figure~\ref{true_vs_claimed_uncertainty}b shows the plot similar to Figure~\ref{true_vs_claimed_uncertainty}a, but for \dnu.  The reported uncertainties are clearly accurate, with typical values of about 2-3\%, similar to what was reported by \citet{Silva20} on about a dozen bright stars.  This shows their results on radius, mass, and age precision are representative for the full sample of red giants observed during 1-2 sectors by TESS when using grid-based modeling including parallax information as performed by \citet{Silva20}.  The figure also illustrates that the red clump stars typically have larger uncertainties (red dashed line) than red giant branch stars, as expected from their more complicated frequency patterns in the power spectra. 

\begin{figure}
\includegraphics[width=\columnwidth]{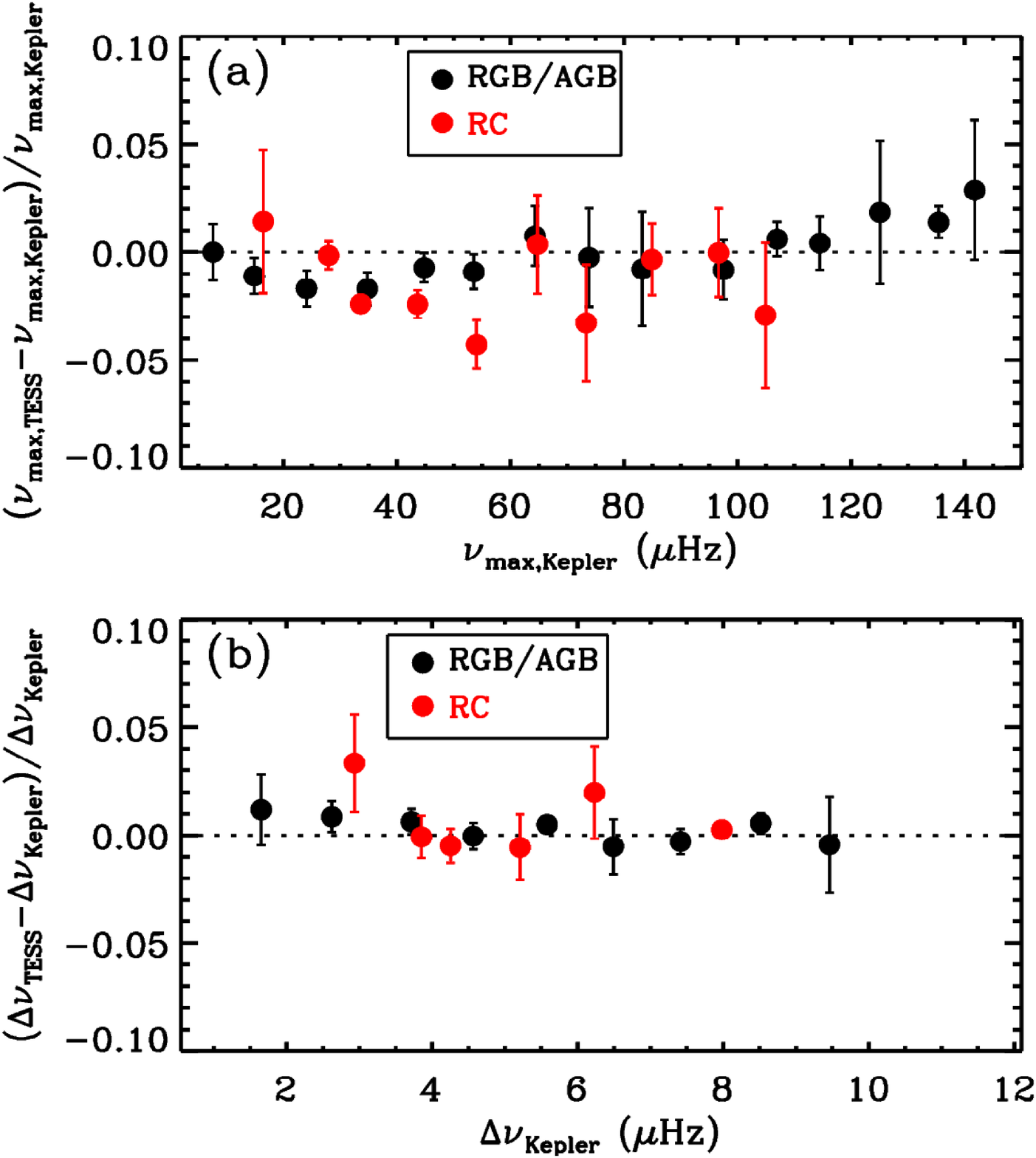}
\caption{Binned fractional difference between TESS and \kepler\ for both \numax\ (a) and \dnu\ (b). Red clump (RC) star identifications are from \citet{Hon18b}.    
\label{tess_kepler_bias}} 
\end{figure}

In addition to random errors, we also want to investigate potential systematics between TESS and \kepler\ results because it can affect comparisons of inferred masses and hence ages of the stars between the two data sets.  Any bias could be either from difference in the data or because the time series are not the same length, which could affect the automated fitting procedures in the data analysis.  In Figure~\ref{tess_kepler_bias} we show the fractional difference between TESS and \kepler\ as a function of \numax\ and \dnu.  Overall there is no strong bias ($-0.004\pm0.003$ for \numax\ and $0.004\pm0.002$ for \dnu).  However, for the red clump (RC) stars with \numax\ around $40\,$\muhz, the TESS \numax\ results tend to be 2-3\% lower than for \kepler.  For red giant branch stars at high \numax\ there is also evidence of some bias (TESS values being larger) but the few data points in these bins makes this somewhat more uncertain.

Finally, we wanted to quantify the scatter in results across the different seismic analysis pipelines that are typically used in large ensembles efforts \citep{Pinsonneault14,Pinsonneault18}. This would act as a way to estimate pipeline-dependent systematic uncertainties in the seismic analysis of TESS data.  
The pipelines were only given the 2724 stars that had confirmed oscillations, and were only asked to provide results deemed reliable.  The pipelines engaged in this analysis were the so-called, A2Z \citep{Mathur10,Garcia14}, BAM \citep{Zinn19b}, BHM \citep{Elsworth17}, CAN \citep{Kallinger10a}, COR \citep{MosserAppourchaux09}, and OCT \citep{Hekker10} updated with packages from TACO (Hekker et al. in prep).  We derived the scatter across pipelines for each star for which at least four pipelines reported a measurement ($\sim 2500$ stars).  
The \numax\ scatter distribution peaked at 2\%, with a slight \numax-dependent trend.  At (low) \numax$=10$-20\muhz\ the scatter was around 3\%, while at (high) \numax$=90$-100\muhz\ the scatter was around 0.5\%.  We found 8\% of the stars had at least one pipeline return an outlier measurement. 
For \dnu, the scatter distribution also peaked at 2\% and with a slight \numax-dependent trend.  At (low) \numax$=10$-20\muhz\ the scatter was around 2-3\%, while at (high) \numax$=90$-100\muhz\ the scatter was around 1-2\%.  About 24\% of the stars had at least one pipeline return an outlier measurement.\footnote{For additional recent details into the biases between pipelines on short time series see for example \citet{Stello17} and \citet{Zinn21b}.} 
These scatter values indicate that the pipeline-dependent systematics are typically smaller than the uncertainties on the individual seismic measurements that are shown Figure~\ref{true_vs_claimed_uncertainty}.

\section{Uncertainty on radius, mass, and age}
Now we turn to the measurement uncertainties on the fundamental stellar properties, which ultimately determine how useful the 1-2 sector red giant data will be for studying the Milky Way.  Based on our results, we expect the minority of the TESS seismic red giant sample will have both \numax\ and \dnu\ measurements available.  We therefore consider two scenarios separately; one where both \numax\ and \dnu\ are available and one where we only have \numax.

We are now in the Gaia era, which gives us powerful additional tools for asteroseismology. Scaling relations are distance-independent, but also highly sensitive to uncertainties in the measurements. The \numax\ and \dnu\ scaling relations give us 
$R \simeq T_{\mathrm{eff}}^{0.5}\nu_{\mathrm{max}}/\Delta\nu^2$ and $M \simeq T_{\mathrm{eff}}^{1.5}\nu_{\mathrm{max}}^3/\Delta\nu^4$ \citep{Brown91,KjeldsenBedding95,Kallinger10}; all variables in terms of solar values, which we assumed to have negligible uncertainty. The utility of a fully asteroseismic approach thus degrades dramatically with increased errors. However, with an independent radius estimate, it is possible to infer masses using either \numax\ or \dnu\ alone. Gaia data can be used to infer luminosity, from a combination of photometry, astrometry, and extinction maps. When combined with spectroscopy, this yields what we will refer to as a Gaia radius.  It is easier to measure \numax\ in TESS data, so following \citet{Stello08} we focus here on \numax\ plus $R$ as an alternative scaling relation. In this case, $M \simeq T_{\mathrm{eff}}^{0.5}\nu_{\mathrm{max}}R_{\mathrm{Gaia}}^2$.  For a sufficiently precise $R$, the uncertainties in the single-seismic-parameter scaling relation can be comparable to, or smaller than, those from two parameter scaling relations.

First, we consider the case where we estimate radii, masses, and ages using seismic scaling relations in the same way as commonly done for \kepler\ and K2 ensemble analyses \citep[e.g.][]{Pinsonneault18,Zinn21b}.  We use standard error propagation for mass uncertainties, and infer the age uncertainty based on the scaling relation from \citet{Bellinger20} (applicable only to red giant branch stars), for which we also needed a typical uncertainty in [Fe/H]. Typical TESS uncertainties are $\sim5\%$ in \numax\ and $\sim3\%$ in \dnu\ (Figure~\ref{true_vs_claimed_uncertainty}).  From the Infrared Flux Method calibrated APOGEE survey, we can expect to be able to obtain \teff\ with uncertainties in the range 40-80K \citep{Casagrande10,Casagrande21}.  We therefore adopt 80K as a conservative random uncertainty value. This leads to typical random uncertainties of 8\% in radius, 19\% in mass, and 63\% in age (the latter assuming an uncertainty in [Fe/H] of 0.1 dex using the `combination 1' formula of table 2 in \citet{Bellinger20}). These uncertainties are dominated by the uncertainties in \numax\ and \dnu.

Systematic errors in \teff\ are $\sim 2$\% \citep{Tayar20}. The 2\% pipeline-to-pipeline scatter in the seimic measurements would add a systimatic of 4\% in radius, 8\% in mass, and 25\% in age. Likewise, the 2-3\% \numax\ bias between TESS and \kepler\ at certain \numax\ ranges (Figure~\ref{tess_kepler_bias}a) translates to a systematic of 2-3\% in radius, 6-9\% in mass, and 20-30\% in age.

Next, we consider the most common scenario where only \numax\ is available. Because the seismic red giant sample from TESS is typically sampling the local neighborhood \citep{Hon21} the stars have relatively small parallax uncertainties from Gaia. For our sample, the median Gaia radius uncertainties are $\sim 6$\% (dominated by parallax uncertainty), assuming photometric temperature inferences and uncertainties, and hence our mass uncertainties are $\sim 12$\%.  This median mass uncertainty, dominated by the radius uncertainty, roughly translates into an expected age uncertainty of 37\% on the red giant branch \citep{Miglio12a}.  In this scenario, the pipeline-to-pipeline systematics and the TESS-to-\kepler\ bias in \numax\, each translate to a systematic of only 2-3\% in mass and 6-9\% in age. \textit{We therefore have the surprising, but robust, result that we can obtain ages with an interesting level of precision using \numax\ alone.}

Spectroscopic \teff\ values will be available for large numbers of survey targets. They are of comparable precision to photometric \teff\ values, but are not subject to systematic errors from extinction (and large metallicity) uncertainties.  APOGEE, for example, is calibrated to be on the Infrared Flux Method scale, with well-controlled random and systematic errors.  Spectra also give powerful composition information, important for inferring ages.  The combination of spectroscopy, Gaia, and \numax\ is therefore likely to be the most fruitful technique for the full TESS asteroseismic sample.
    
Finally, we note that these estimates assume the `typical' results (the median of the uncertainty distribution), and the error model is based on only 1-2 sectors of data.  Clearly, the best fraction of stars will provide significant lower radius, mass, and age uncertainties. As an example, for the closest stars, parallax uncertainties are smaller and hence the uncertainties in the bolometric corrections and \teff\ (including the 2\% \teff\ systematic error) will dominate the radius error budget. Considering only these stars, we would expect internal median radius uncertainties of 3-4\%, mass uncertainties of 8-9\%, and hence about 25-30\% in age, even when only \numax\ is measured.  Also, the use of grid-based modelling (adding isochrone constraints on stellar inferences) will improve results as demonstrated by \citet{Silva20}, who achieved $\sim 3$\% in radius, $\sim 6$\% in mass, and $\sim 20$\% in age (internal uncertainties) when including Gaia parallaxes for stars with \numax\ and \dnu\ uncertainties similar to our sample.  Uncertainties will of course also be lower for stars with longer time series \citep{Hekker12}, which will be achieved with the ongoing extended TESS mission, especially in the continuous viewing zones \citep{Mackereth21}.  Photometry optimized for asteroseismology \citep{Handberg21,Lund21} is also expected to lead to lower uncertainties and larger detection yields.

\section{Confusion from blends}
TESS has relatively large pixels (21 arcsec on sky) compared to \kepler\ (3.98 arcsec) and blending is therefore expected to be more common with TESS.  Blending can dilute the signal of target stars and hence lower detection yields \citep[][Fig. 3]{Mackereth21}. In addition, blends can cause `confusion', where the seismic signal from one star is imposed on that of another.  
Among our red giant targets we noticed this confusion when identifying the seismic detections. It manifested as two nearby stars showing almost identical power spectra, dominated by the star with highest amplitude oscillations. In our case, confussion could only occur between red giants because less evolved stars oscillate at frequencies above the Nyquist frequency and with amplitudes too low to cause confusion \citep{GarciaStello15}. To quantify how common confusion would typically be in a sample like ours, we applied a similarity measure on power spectra displayed in units of power density versus log frequency.  The similarity that we used is known as the Shape-Based Distance \citep{PaparrizosGravano15}.  It quantifies the correlation between two arrays, $x$ and $y$, as $CC(x,y)/(\| x\|\| y\|)$, where  $CC$ is the cross-correlation operator and $\| \cdot \|$ indicates the vector norm. The Shape-Based Distance has a value of zero for a perfect correlation and $-1$ for a perfect anti-correlation.  Before calculating the Shape-Based Distance between the TESS power spectra of two stars, we first bin each spectrum (in log units) into an array of length 1000.  Next, we applied Gaussian smoothing with a kernel size of 15 to the binned spectrum and normalized the spectrum to have a mean value of zero and a standard deviation of one.

For each target star, we identified another star within our sample that has the smallest Shape-Based Distance.  If this other star had an angular separation less than the typical photometric aperture to the target star (150 arcseconds), it was flagged as a potential nearby blending star.  To further vet blending star candidates, we ensured that the power was the same within the oscillation power excess for a target star and its candidate blending star.  Using the binned and smoothed spectrum, we calculated the mean difference in power within the Full-Width-Half-Maximum of the oscillation power excess \citep[$\delta\nu_{\mathrm{FWHM}}=0.59\nu_{\mathrm{max}}^{0.90}$;][]{Mosser10} between a target star and its blending companion.  This power excess difference should be small for a correctly identified blending star compared to that of any other star that is not the true source of the blending.  Therefore, each blending candidate was verified to be a blend, only if its power excess difference puts it in the top 0.5\% percentile of most similar excesses compared to those of all other stars in our sample.  This vetting process combining Shape-Based Distance and power differences near \numax, effectively identified blends that have power spectra that are very similar to a target star.

\begin{figure*}
\includegraphics[width=17.8cm]{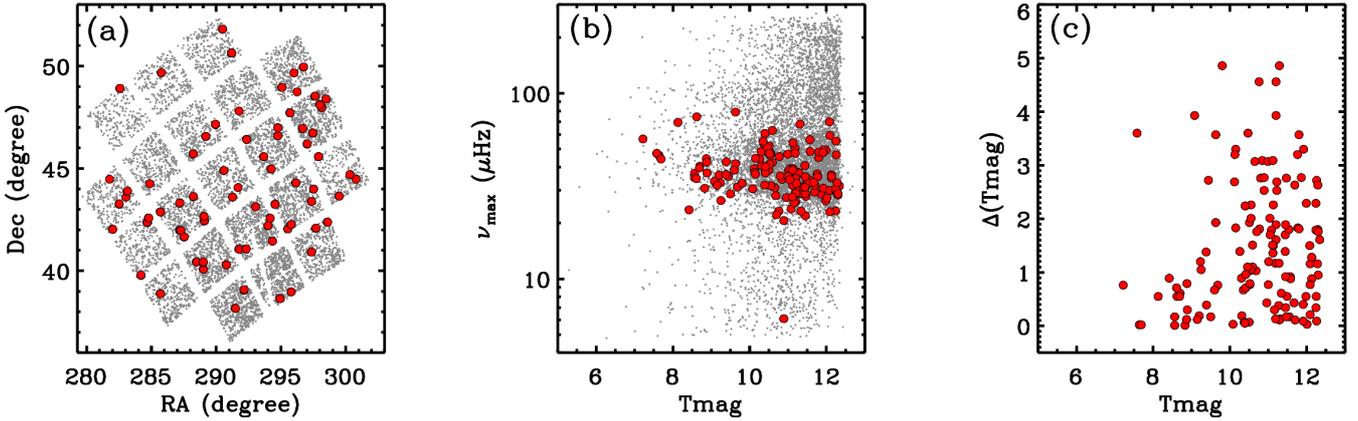}
\caption{Location of blends (red dots) on the sky (panel a) and in the \numax-Tmag plane (panel b). Grey points are all stars in our sample as in Figures~\ref{fov} and \ref{numax_vs_tmag}. (c) Magnitude difference between target and blending star as function of the target's magnitude.    
\label{blends}} 
\end{figure*}

A total of 85 targets, or about 1\% of our red giant sample, were found to be confused due to blending (counting any pair of blends only once).  These stars did not show up as a particularly discrepant set in the previous figures.  Figure~\ref{blends}a shows the sky position of these blends, while Figure~\ref{blends}b shows their location in the \numax-Tmag plane.  In Figure~\ref{blends}c we show the difference in magnitudes between target and blending star as a function of the target's magnitude.
So for seismic ensemble analyses of field red giants with TESS brighter than Tmag of 12.5, confusion due to blending is a relatively minor issue.  Towards fainter magnitudes (and in particularly crowded fields), the issue will of course be more severe.  However, Figure~\ref{numax_vs_tmag} shows that we can only expect to detect oscillations in fainter stars if they are quite luminous, which comprises a small fraction of all red giants that TESS will be able to detect oscillations in.

\section{Conclusion}
Our findings, based on 1-2 sectors of TESS data, can be summarised as the following:
\begin{itemize}
 \item Due to photon noise, oscillations are typically not detectable in low luminosity red giant stars (\numax\ $\gtrsim 150\,$\muhz; $\log g\gtrsim 3.1\,$dex) except for the brightest stars (Tmag $\lesssim 8-9$). This is in agreement with \citet[their Fig. 10]{Mosser19}. 
 \item Our results suggest TESS will be able to detect oscillations down to Tmag$\sim$14 for the most luminous giants (\numax\ $\lesssim$ 10\muhz; $\log g\lesssim 1.9$).
 \item Of the stars with detected oscillations we can measure \dnu\ reliably in about 20\% of them, but this yield depends a lot on the type of star (its \numax\ and if it is He-core burning or not) and the amount of TESS data available.
 \item We find the median random uncertainty is 5-6\% for \numax\ and 2-3\% for \dnu, which for common grid-modelling approaches should yield uncertainties of 3\% in radius, 6\% in mass, and 20\% in age \citep{Silva20}.
 \item For stars with only a \numax\ measurement -- the most common case for TESS -- we obtain median uncertainties of 6\% in radius and 12\% in mass (hence expected 37\% in age) based on the \numax\ scaling relation and Gaia parallax measurements.
 \item Systematics in the \teff\ scale, pipeline-to-pipeline scatter in the seismic results, and bias between TESS and \kepler\ results each translate to systematics of 2-3\% in radius, 6-9\% in mass, and 20-30\% in age. 
 \item Our blending analysis of the \kepler\ field, which sits between Galactic latitudes of 6 and 21 degrees, suggest confusion of seismic signals from neighboring stars due to blending is not expected to affect more than 1\% of red giants observed by TESS. 
\end{itemize}
Finally, we note that this investigation is based on a single set of light curves.  It would be desirable in future to quantify detection yields from independent asteroseismic-optimised light curves when they become available in the \kepler\ field such as the forthcoming TASOC light curves (Handberg et al. in press; Lund et al. in press).

\section*{Acknowledgments}
We thank Kosmas Gazeas for comments on the manuscript.
D.S. is supported by the Australian Research Council (DP190100666).
N.S. and D.H. acknowledge support the National Aeronautics and Space Administration (80NSSC18K1585, 80NSSC20K0593) awarded through the TESS Guest Investigator Program. D.H. also acknowledges support from the Alfred P. Sloan Foundation. 
T.R.B. is supported by the Australian Research Council (DP210103119).
R.A.G. acknowledges the support from the PLATO CNES grant.
S.M. acknowledges support by the Spanish Ministry of Science and Innovation with the Ramon y Cajal fellowship number RYC-2015-17697 and the grant number PID2019-107187GB-I00.

\section*{Data availability}
The  data  underlying  this  article  are  available  upon  request.




\bibliographystyle{mnras}
\bibliography{bib_complete} 








\bsp	
\label{lastpage}
\end{document}